\documentclass[11pt]{article}
\usepackage{graphicx,epsfig}

\newcommand{\BABARPubYear}    {01}

\newcommand{\BABARProcNumber} {70}
\newcommand{\SLACPubNumber} {9037}

\input babarsym
\newcommand\prd[3]   {{{Phys.\ Rev.\ }{\bf D #1} (#2) #3}}

\newcommand\prl[3]   {{{Phys.\ Rev.\ Lett.\ }{\bf #1} (#2) #3}}

\renewcommand{\hepex}[1]{{hep-ex/#1}}

\newcommand\jnim[3]   {{{\it Nucl.\ Instrum.\ Meth.\ }{\bf #1} (#2) #3}}
\newcommand\jplb[3]   {{{\it Phys.\ Lett.\ }{\bf B #1} (#2) #3}}
\newcommand\jepjc[3]  {{{\it Eur.\ Phys.\ J. }{\bf C #1} (#2) #3}}

\long\def\inst#1{\par\nobreak\kern 4pt\nobreak
    {\it #1}\par\vskip 10pt plus 3pt minus 3pt}

\begin{document}
{\pagestyle{empty}

\begin{flushright}
SLAC-PUB-\SLACPubNumber \\
\babar-PROC-\BABARPubYear/\BABARProcNumber \\
\hepex 0204001 \\
November, 2001 \\
\end{flushright}

\par\vskip 4cm

\begin{center}
\Large \bf Inclusive Semileptonic $B$ Decays at BABAR
\end{center}
\bigskip

\begin{center}
\large 
U. Langenegger\\
Stanford Linear Accelerator Center \\
Stanford University, Stanford, CA 94309, USA \\
(for the \lbabar\ Collaboration)
\end{center}
\bigskip \bigskip

\begin{center}
\large \bf Abstract
\end{center}
We  present  measurements   of  the  inclusive  semileptonic
branching  fractions   of  charged   and  neutral  $B$   mesons  using
$20.6\invfb$ of data measured at the $\FourS$ with the
\babar\  detector.   Events  are  tagged with  a  fully  reconstructed
hadronic decay of a $B$  meson.  The correlation between the flavor of
the tag  $B$ meson  and the electron  charge allows the  separation of
prompt semileptonic $B$ decays and cascade semileptonic charm decays.
We obtain the preliminary inclusive semileptonic branching fraction of
charged $B$  mesons $b_+ = 0.103\pm0.006\pm0.005$,  neutral $B$ mesons
$b_0     =    0.104\pm0.008\pm0.005$,     their    average     $b    =
0.104\pm0.005\pm0.004$,     and     their     ratio     $b_+/b_0     =
0.99\pm0.10\pm0.03$.

\vfill
\begin{center}
Contributed to the Proceedings of the International 
Europhysics Conference on High Energy Physics (HEP 2001), \\
7/13/2001 --- 7/18/2001, Budapest, Hungary
\end{center}

\vspace{1.0cm}
\begin{center}
{\em Stanford Linear Accelerator Center, Stanford University, 
Stanford, CA 94309} \\ \vspace{0.1cm}\hrule\vspace{0.1cm}
Work supported in part by Department of Energy contract DE-AC03-76SF00515.
\end{center}

\section{Introduction}
In  this   paper  we  report   a  preliminary  determination   of  the
semileptonic  branching  fractions  $b_+$  and $b_0$  of  charged  and
neutral  $B$  mesons,  respectively,  obtained  with  measurements  of
electron\footnote{Charge-conjugation is implied throughout this note.}
spectra in events tagged by  a fully reconstructed hadronic decay of a
$B$ meson.  Electrons from the prompt semileptonic decay of the second
$B$ meson  are separated from electrons of  cascade semileptonic charm
decays  using the  correlation  between the  electron  charge and  the
flavor of the tag $B$ meson.


\section{Data sample and detector}
The data for this analysis were collected with the \babar\ detector at
the PEP-II asymmetric $e^+e^-$ collider \cite{ref:PEPII} in 1999--2000
and amount  to an integrated  luminosity of $20.6\invfb$ taken  at the
$\Upsilon(4S)$.

The  \babar\ detector  is a  large-acceptance  solenoidal spectrometer
\cite{ref:BABAR}.  All components  of  the detector  are  used in  the
present analysis.
Charged  particles are detected  and their  momenta are  measured with
high  precision in  a five-layer  double-sided silicon  vertex tracker
(SVT)  and a  40-layer drift  chamber (DCH).   Both SVT  and  DCH also
provide   ionization    measurements   used   in    charged   particle
identification.     Additional    charged   particle    identification
information  is obtained  from the  Cherenkov angle  measurement  in a
detector of internally reflected Cherenkov light (DIRC).
Electromagnetic  showers  of   photons  and
electrons   are   measured    with   excellent   resolution   in   the
electromagnetic   calorimeter   (EMC),   composed  of   6580   CsI(Tl)
crystals. The  iron flux return  (IFR), used in the  identification of
muons, is segmented and instrumented with multiple layers of resistive
plate chambers.

\section{Event selection}

We  select  multihadron  events  containing  one  fully  reconstructed
hadronic  decay  of  a $\Bz$  or  $\Bp$  candidate  and at  least  one
additional   charged  track.    Continuum  background   consisting  of
non-resonant $e^+e^-\to  q\overline{q}$ production $(q = u,  d, s, c)$
is  reduced  by  requiring the  ratio  of  the  second to  the  zeroth
Fox-Wolfram moment \cite{ref:fw} to be $R_2 < 0.5$.

Electron candidates must  have a ratio of calorimeter  energy to track
momentum of $0.89 < E/p <  1.2$.  The profile of the energy deposit is
required  to be  consistent with  an electromagnetic  shower,  both in
lateral \cite{ref:lat}  and azimuthal \cite{ref:zernike}  extent.  The
ionization measurement in the DCH  and the Cherenkov angle measured in
the  DIRC  (if available)  are  required  to  be consistent  with  the
electron hypothesis  at the $3\sigma$ level.  Candidates  that are not
matched to an EMC cluster are retained if their $dE/dx$ measurement in
the   DCH   is  consistent   with   the   electron  hypothesis.    The
misidentification probability for hadrons faking an electron signature
in the detector  is measured in data with samples  of pions, kaons and
protons that have been  identified by kinematic requirements with high
purity.

Muon candidates  must pass requirements  on the measured  and expected
number of  interaction lengths penetrated, the  position match between
the extrapolated track and IFR hits,  and on the average and spread of
the number of IFR hits per layer.  

Kaon candidates  are identified with $dE/dx$ measurements  both in the
SVT and  DCH and the number  of Cherenkov photons  and Cherenkov angle
measured in the DIRC.

A photon candidate  consists of a localized energy  maximum in the EMC
with a  lateral energy profile consistent  with a photon  shower and a
minimum energy of $E > 30  \mev$. Pairs of photons measured in the EMC
are constrained  to the known $\piz$  mass if their  invariant mass is
$115 < m_{\gamma\gamma}  < 150 \mev$ and if the  sum of their energies
is larger than $100\mev$.

Pairs of charged pions are  combined to form $\KS$ candidates if their
invariant  mass is  between 462  and $534\mev$.   Neutral  and charged
kaons are  combined with pions  to form $\Kstar$ candidates,  if their
invariant  mass  is  within  $100\mev$  of  the  nominal  $K^*$  mass.
$\rho^+$ and $\rho^0$ candidates  are reconstructed in the decay modes
$\pip\piz$  and   $\pip\pim$,  if  their  invariant   mass  is  within
$\pm150\mev$  of  the nominal  $\rho$  mass.   $a_1^+$ candidates  are
formed by  combining three charged  pions, if their invariant  mass is
between $1.0$ and  $1.6\gev$ and the $\chi^2$ probability  of a vertex
fit of the $a_1^+$ candidate is greater than $0.1\%$.

$\Dzb$ candidates  are reconstructed in the  decay channels $\Kp\pim$,
$\Kp\pim\piz$, $\Kp\pip\pim\pim$,  and $\KS\pip\pim$. $\Dm$ candidates
are formed in the $\Kp\pim\pim$ and $\KS\pim$ decay modes.  
$\Dstarm$ candidates are reconstructed by combining a $\Dzb$ candidate
and a  soft pion. The  selection of $\Dstarm$ candidates  requires the
mass  difference $\Delta  m =  m_{\Dzb\pim} -  m_{\Dzb}$ to  be within
$\pm3\sigma$   of  the  nominal   value.   $\Dstarz$   candidates  are
reconstructed  by combining  a  $\Dz$  candidate with  a  $\piz$ or  a
photon.  $\jpsi$  candidates  are  reconstructed in  the  decay  modes
$e^+e^-$ and  $\mu^+\mu^-$ where at  least one decay daughter  must be
positively identified as a lepton.
$\jpsi$ candidates must have an invariant mass of $2.95 < m_{e^+e^-} <
3.14  \gev$ and  $3.06 <  m_{\mu^+\mu^-} <  3.14  \gev$, respectively.
$\psi(2S)$  candidates   are  reconstructed  in   the  decay  channels
$\psi(2S) \to e^+e^-, \mu^+\mu^-$ and $\psi(2S) \to J/\psi \pip\pim$.

$\Bz$ candidates are formed by combining a $D^{(*)-}$ candidate with a
$\pip$, $\rho^+$,  or $a_1^+$.  In addition, we  reconstruct the decay
channels $\Bz\to\jpsi K^{(*)0}$  and $\Bz\to\psi(2S) K^{(*)0}$.  $\Bp$
candidates are reconstructed by  combining a $D^{(*)0}$ candidate with
a  $\pip$,  $\rho^+$, or  $a_1^+$.   Furthermore,  the decay  channels
$\Bp\to\jpsi    K^{(*)+}$    and    $\Bp\to\psi(2S)   K^{(*)+}$    are
reconstructed.

The selection  of $B$ candidates  is based on two  nearly uncorrelated
variables, the  difference $\Delta  E$ between the  energy of  the $B$
candidate and  the beam energy  in the $\FourS$  center-of-mass frame,
and  the beam-energy  substituted  mass $m_{ES}  =  \sqrt{(s/2 +  {\bf
p}\cdot  {\bf p}_i)^2/E_i^2  - {\bf  p}^2},$ where  $\sqrt{s}$  is the
center-of-mass energy, $E_i$ and ${\bf  p}_i$ are the total energy and
the momentum of  the initial state in the  laboratory frame, and ${\bf
p}$ is  the momentum of the $B$  candidate in the same  frame.  In the
$(m_{ES}, \Delta  E)$ plane a mode-dependent signal  region is defined
as  $|m_{ES} -  m_{ES}^{peak}|  < 3\sigma(m_{ES})$  and  $|\Delta E  -
\Delta   E^{peak}|  <  3\sigma(\Delta   E)$,  where   the  resolutions
$\sigma(m_{ES})$  and $\sigma(\Delta  E)$  are measured  in data.   If
several $B$  candidates fulfill these requirements, only  the one with
the smallest $|\Delta  E - \Delta E^{peak}|$ is  retained.  Signal and
background   yields  are   extracted   from  fits   to  the   $m_{ES}$
distributions of  $B$ candidates with a Gaussian  distribution for the
signal and a background given by \cite{ref:argus}

$$
{dN\over dm_{ES}} \propto m_{ES} \times 
\sqrt{1 - {m_{ES}^2\over E^{*2}_{beam}}}\times
\exp\left[-\zeta\left(1 - {m_{ES}^2\over E^{*2}_{beam}}\right)\right],
$$

\noindent  where  the  only   free  parameters  are  $\zeta$  and  the
normalization.   In  total,  we  reconstruct  $7684\pm120_{stat}\,  \Bp$
candidates and $6533 \pm112_{stat}\, \Bz$ candidates.

\section{Measurement of the inclusive semileptonic branching fractions}
In  events  with a  reconstructed  tag  $B$  candidate we  search  for
electron candidates  with a well  measured track originating  from the
known beam-spot.  Requirements on the momentum $p > 0.5 \gev$ (both in
the  laboratory  and  center-of-mass   frames)  and  the  polar  angle
$23^\circ < \theta < 135.9^\circ$  in the laboratory frame are applied
to ensure good discrimination against hadrons and the full containment
of the shower in the  EMC. The mean electron identification efficiency
is $(89.8\pm0.7_{stat})\%$.   

This   electron   spectrum  is   corrected   for  various   background
contributions  as  described below.   Electrons  from $\jpsi$  decays,
photon conversions in the beam  pipe and support structure of the SVT,
and $\piz\to e^+e^-\gamma$ are detected using the invariant mass of an
electron  paired with  a track  of  opposite charge.   We correct  the
background spectrum for the inefficiency of the pair-finding algorithm
and subtract the resulting  spectrum from the total electron spectrum.
The contribution  of electrons from  converted photons is  sizeable in
the lowest momentum bin, while electrons from $\jpsi$ decays make only
a  very small  contribution at  high momenta.   The  contribution from
continuum events and combinatoric  background in the reconstruction of
the  tag $B$ meson  is obtained  with $B$  candidates in  the $m_{ES}$
sideband and  amounts to  $4.2\%$ of the  total spectrum.   The hadron
fake contribution  is estimated using  misidentification probabilities
for  all hadronic  tracks  in  tagged events.   The  mean hadron  fake
probability  is $(0.25\pm0.005_{stat})\%$ and  is dominated  by kaons.
The fake contribution is largest at low momenta and amounts to $4.3\%$
of the total electron spectrum.

Various processes other than prompt semileptonic $B$ decays contribute
to the  electron spectrum as  well.  Electrons from $\tau$  decays are
estimated  by Monte Carlo  simulation, where  a branching  fraction of
${\cal B}(B\to  \tau X)  = 2.6 \pm  0.4\%$ is  assumed \cite{ref:pdg}.
Electrons from  charm mesons originate  either from $b\to  c$ (``lower
vertex''  charm)  or  from  $W^-\to \overline{c}s$  (``upper  vertex''
charm).  The latter have the  same charge as electrons from prompt $B$
decays and constitute  a background, while the former  are of opposite
charge and are accounted for with the charge correlation to the flavor
of the  tag $B$ meson.   Electrons from semileptonic  ``upper vertex''
charm decays are estimated by  Monte Carlo simulation, where the world
average  of the  branching fractions  for $B\to  D_sX$,  $B\to D^{(*)}
\overline{D}^{(*)}K^{(*)}$,   $D_s\to  eX$,   $D_s\to  \tau   X$,  and
semileptonic $D^0$  and $D^+$  decays have been  assumed. Semileptonic
$B$ decays  in the  \babar\ Monte Carlo  simulation have  been modeled
according to the charm meson  involved. A parametrization of HQET form
factors, defined  in \cite{ref:duboscq},  is used for  $B\to D^*e\nu$,
the model of Goity and Roberts \cite{ref:goityroberts} is used for the
non-resonant decays  $B\to D^{(*)} \pi e\nu$, whereas  the ISGW2 model
\cite{ref:isgwtwo} is used for all other semileptonic decays.


\begin{figure}
\begin{center}\epsfig{file=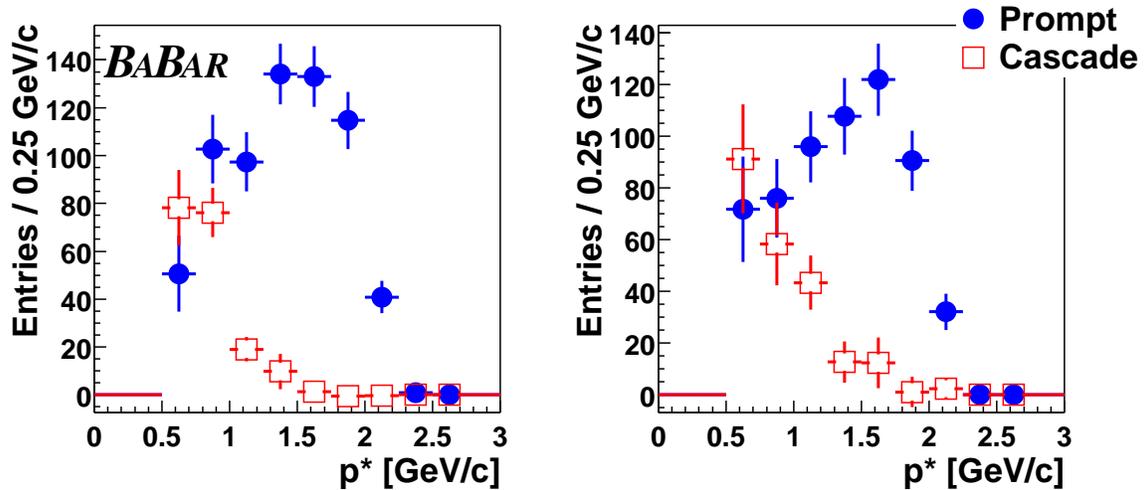,width=15cm}
\end{center}
        \caption{Measured
        electron spectra  from $B\to Xe\nu$ (filled  circles) and from
        $b\to  c\to Ye\nu$  (open squares)  measured in  events tagged
        with $\Bp$ (left) and $\Bz$ (right) mesons.}
	\label{fig:espectrum}
\end{figure}

The correlation between  the charge of the electron  and the flavor of
the reconstructed tag $B$ meson  allows the separation of the measured
electron  spectrum  into contributions  from  prompt semileptonic  $B$
decays and  from cascade semileptonic  charm decays (from  the ``lower
vertex''). For events tagged with  a $\Bz$ meson, $\BB$ mixing must be
taken into account with
%
%
the   $B_d$   mixing   probability   $\chi_d  =   0.174   \pm   0.009$
\cite{ref:pdg}.  In  events tagged with a $\Bp$  $(\Bz)$ candidate, we
find $674\pm34_{stat}$ ($597\pm38_{stat}$) prompt electrons.  Figure~1
shows  the spectra  measured in  events  tagged with  $\Bp$ and  $\Bz$
candidates.   The  prompt  $B$  semileptonic branching  fractions  are
derived from these numbers and  the normalization (given by the number
of reconstructed  $B$ mesons) after correcting for  (1) the difference
in  event selection efficiency  between events  where both  $B$ mesons
decay  fully  hadronically  and  events  where one  $B$  meson  decays
semileptonically%
, (2) the geometric acceptance  of electron detection of $86.2\%$, and
(3) the extrapolation of the momentum spectrum of $(5.3\pm0.2)\%$.

The measured  values of  the branching fractions  are for  charged $B$
mesons $b_+ = 0.103 \pm0.006 \pm0.005$ and for neutral $B$ mesons $b_0
= 0.104 \pm0.008  \pm0.005$.  The ratio of the  branching fractions is
determined to be $b_+/b_0 = 0.99 \pm0.10 \pm0.03$.
These  results are  consistent  with and  comparable  in precision  to
\cite{ref:cleocharge}.   The  average branching  fraction  $b =  0.104
\pm0.005 \pm0.004$  is consistent with the many  other measurements at
the $\FourS$ and LEP.

The  systematic errors  are  dominated by  three  sources: First,  the
uncertainty in  the track reconstruction efficiency  together with the
systematics of  the cross-feed in  the $B$-reconstruction amount  to a
total of  $3.5\%$.  Second, electron identification  ($1.8\%$) and the
uncertainty  in   the  hadron   fake  subtraction  ($1.4\%$)   have  a
significant  contribution to  the electron  spectrum.   The systematic
errors associated with pair backgrounds from converted photons, Dalitz
decays  of $\piz$,  and  $\jpsi$ decays  are  negligible.  Third,  the
uncertainty of physics processes contributing to the electron spectrum
(mostly from upper vertex charm  decays) amounts to $1.7\%$.  Only the
first  group of  errors  affects  the determination  of  the ratio  of
branching   fractions,  the   other  two   groups  of   errors  cancel
approximately.

\section{Conclusions} 
We  have  presented  a   measurement  of  the  inclusive  semileptonic
branching fractions of charged and neutral $B$ mesons in events tagged
with  a fully  reconstructed  hadronic  decay of  a  $B$ meson.   This
analysis is statistics limited  but already of comparable precision to
the current world  average. The systematic error can  be improved with
more  statistics, leaving  room for  significant improvements  as more
data are accumulated at \babar.

\end{document}